\input{aipcheck}

\documentclass[
    ,final            
  ]
  {aipproc}

\layoutstyle{8x11double}

\begin{document}

\title{Eccentric Binary Millisecond Pulsars}


\author{Paulo C. C. Freire}{
  address={Arecibo Observatory, HC 3 Box 53995, Arecibo PR 00612, USA}
  ,altaddress={West Virginia University, PO Box 6315, Morgantown WV 26505, USA} 
}





\begin{abstract}
In this paper we review recent discovery of millisecond pulsars
(MSPs) in eccentric binary systems. Timing these MSPs we were able to
estimate (and in one case precisely measure) their masses. These results
suggest that, as a class, MSPs have a much wider range
of masses (1.3 to $> 2 M_{\odot}$) than the normal and mildly recycled pulsars
found in double neutron
star (DNS) systems ($1.25 < M_p < 1.44 M_{\odot}$). This is very likely to be
due to the prolonged accretion episode that is thought to be required to
form a MSP. The likely existence of massive MSPs makes them a powerful probe
for understanding the behavior of matter at densities larger than that of the
atomic nucleus; in particular, the precise
measurement of the mass of PSR J1903+0327 ($1.67 \pm 0.01 M_{\odot}$) excludes
several "soft" equations of state for dense matter. 
\end{abstract}

\maketitle


\section{Introduction}

In recent years, more than a dozen MSPs have been found in binary systems
with eccentric orbits. These new and unexpected discoveries will
allow precise measurements of the masses of several MSPs. The aim of this
review is to convey the reasons why this is an exciting development.

In the first two sections, we will provide the context for this work.
We describe how binary pulsars form; this will allow us to understand
why they are found in two main distinct groups with different
spin and orbital characteristics.
We then describe briefly how we measure neutron star (NS)
masses (and, in a few cases test general relativity, GR) with radio
timing.

In the third section, we enter the core of the review, i.e., we describe
the recent discovery of many MSPs in eccentric
binaries. Most of these are located in globular clusters.
We also review the properties of an intriguing new system, PSR~J1903+0327,
the first eccentric binary MSP in the galactic disk.
We finally review what can be learned from these new binary systems.

We assume that the reader is familiar with the concept of 
NSs, and has an idea of what pulsar is: a NS with strong, anisotropic
electromagnetic emission. For distant observers, this emission is modulated
by the rotation of the object in a repeatable, clock-like fashion,
much like the light ``pulses'' of a lighthouse. In what follows,
we concern ourselves with radio pulsars only.

\section{Formation of Binary Pulsars}
\label{sec:binaries}

Of a total of 1826 rotational-powered radio pulsars listed in the
ATNF catalog \cite{mhth05}, 141 are found in binary systems,
which makes them relatively rare.
Their formation is described in Lorimer
(2008)\nocite{lor08} and references therein. Fig. 1 is used there
to summarize the two main channels for the formation of these systems.

\begin{figure}
  \includegraphics[height=0.54\textheight]{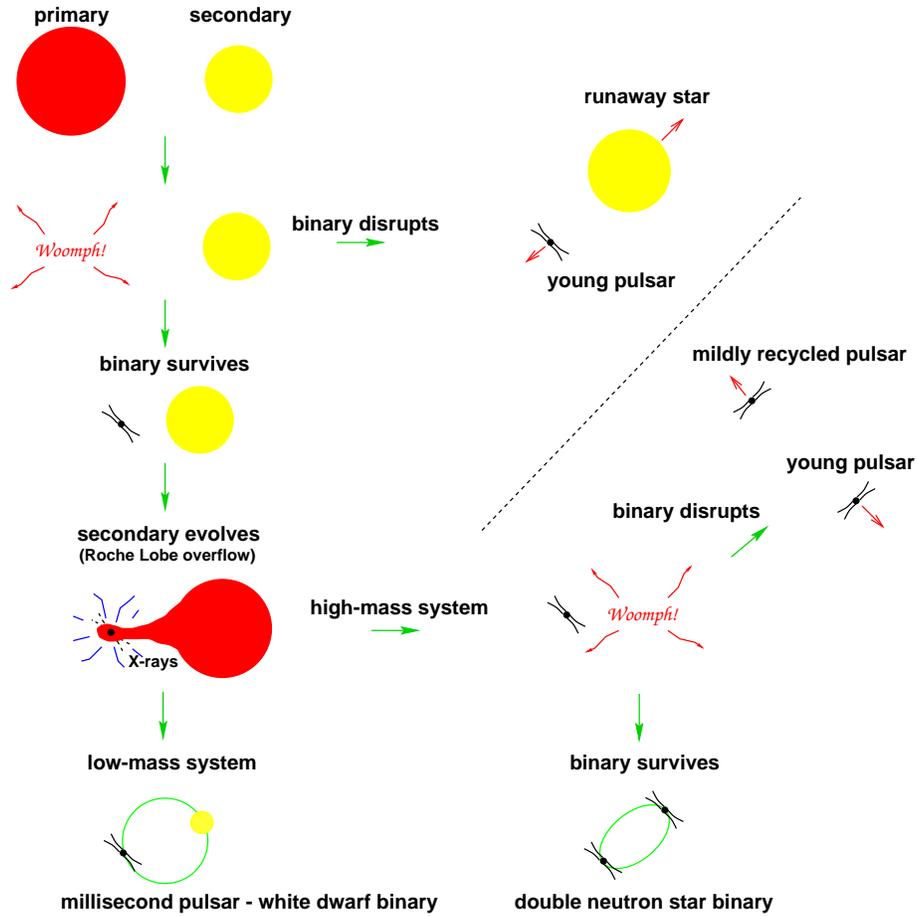}
  \caption{Formation of millisecond pulsars. From Lorimer (2008).}
\end{figure}

At the top left, we start with a binary system with two
main-sequence (MS) stars where at least one of the
components has $M_{*} > 8 M_{\odot}$. Such stars are bound to end their lives
as supernovae (SN) after only a few Myr of evolution (next below). This can
disrupt the binary system, although the exact probabilities depend on the
orbital parameters, the previous masses of the binary components and
the magnitude and direction of the SN kick.

If the system survives the first SN, we observe a NS
orbiting a MS star. There are at least four known examples of
binaries at this particular evolutionary stage where the NS is
observed as a radio pulsar. As the companion evolves, it eventually
fills its Roche lobe\footnote{The Roche lobe is the region around
a member of a binary system where matter can still be
gravitationally bound to it.}. At this stage, transfer of matter from
the companion to the NS creates an ``X-ray binary''\footnote{X-ray emission
is generated by the hot gas in the accretion disk around the NS, in the direct
impact of matter with the surface of the NS
and in some cases in thermonuclear deflagration of matter accumulated at
its surface, giving X-ray binaries a rich observational
phenomenology. High-energy emission (in X and gamma-rays) is also generated
by the interaction of the pulsar wind with the companion's wind.}.
The main consequence for the neutron star is spin-up: the accretion
of matter from the companion transfers angular momentum from the orbit to the
NS. This can make the NS reappear as a radio pulsar, in which case it is
described as a "recycled" pulsar.

The nature of these X-ray binaries and what happens during subsequent
evolution is determined mainly by the mass of the companion.

\begin{enumerate}

\item If the companion's initial mass is also above the $8 M_{\odot}$ threshold,
then it is also fated to explode as a SN and form a second NS.
If the SN kick is correctly aligned, the two NSs remain bound, forming a
DNS system. One of the NSs might then be detected as
a recycled radio pulsar, as in the case of PSR~B1913+16 \cite{ht75}, and
the other might be detected as a young radio
pulsar, as in the case of PSR~J1906+0746 \cite{lsf+06}. In one known case,
PSR~J0737$-$3039, both NSs are detectable as radio pulsars \cite{lbk+04}.

These systems always have {\em eccentric orbits}: the NSs
{\em behave like point masses}, so no tidal circularization happens after the
second SN. As we will see below, these two features make these systems 
especially useful in the study of gravitation.

\item If the companion's mass initial mass is below the $8 M_{\odot}$
threshold the companion will evolve much more slowly and eventually
form a white dwarf (WD) star \cite{acrs82}. Because
the companion is an extended object for up to several Gyr
after the first SN, the orbit is very likely to be
tidally circularized (unless the separation between components is large).
No SN or other sudden events occur later on that can change this
state of affairs. For that reason, the vast majority of pulsar - WD systems
found in the disk of the Galaxy (and all those with orbital periods
smaller than a few hundreds days) have nearly circular orbits.

\end{enumerate}

The different accretion histories have other important implications, which
matter directly to the work described here:

\paragraph{Spin Periods}

If the companion is light and forms a WD its evolution timescale is of the order
of several Gyr. The long accretion episodes made possible by this
slow evolution result in NS spin frequencies of several hundred Hz, i.e,
the NS becomes a ``millisecond pulsar'' (MSP). The first
example of this class, PSR~B1937+21, was discovered in 1982
\cite{bkh+82}. With a spin frequency of 642 Hz, this was
until 2006 the fastest spinning pulsar known.

If the companion is massive and forms a NS its evolutionary
timescale is of the order of a few Myr. Given the much shorter accretion
episode, the resulting spin frequencies for the accreting NS are one order
of magnitude smaller than for MSPs.
For that reason, we designate them here as ``centi-second pulsars'' (CSPs).

\paragraph{Magnetic fields}

One apparent consequence of accretion is a greatly diminished
surface magnetic dipole. CSPs, and to a greater extent MSPs have magnetic
dipoles 3 to 5 orders of magnitude smaller than those of normal
pulsars. This results in a much smaller braking torque and a much longer
lifespan (of the order of a Hubble time) as radio pulsars.

Another very important
effect of the much smaller torque is a much cleaner rotation: effects like
``timing noise'' and ``glitches'', which make the rotational phase of
most young pulsars unpredictable, is greatly diminished or even absent in
CSPs and MSPs. This is fortunate, because the short spin periods
allow very precise radio monitoring of the spin phase of these pulsars.
This monitoring shows that some of the latter objects appear to be more stable than 
atomic clocks; this makes them extremely useful astrophysical tools
(see below).

\paragraph{Masses}

Another consequence of the different accretion histories has to do
with mass transfer. Since the accretion episode necessary for the formation
of a MSP is much longer than for CSPs, one can reasonably expect that the
amounts of matter accreted by the former are much larger than
by the latter. However, it is not clear whether this should introduce
a systematic difference in mass between CSPs and MSPs. The recent results
on MSP mass measurements (see below) shed some light on this matter.

\section{How to Measure the Mass of a Neutron Star}
\label{sec:measuring}

As we described above, among radio pulsars recycled pulsars have
the highest timing precision and the most stable rotation. Furthermore,
a majority of them are found in binary systems, where such features
can be more profitably employed.

After discovery of a binary pulsar, we can use Doppler variations in
the spin period to measure the orbital velocity changes along the
line-of-sight. The situation is analogous to that of
spectroscopic binaries, where the changing Doppler shift continuously
changes the wavelengths of the star's spectral lines. Because we can only
measure changes of velocity along the line of sight our knowledge of the
binary parameters is incomplete.

\subsection{Keplerian Orbits}

In the case of a Keplerian orbit, the measured variations of the
line-of-sight velocity can lead (after a least-squares fit to the observations)
to a determination of five "Keplerian" parameters: the
orbital period ($P_b$), the semi-major axis of the orbit projected along the
line of sight ($x$; for pulsars, this quantity is normally expressed in
light seconds), the orbital eccentricity ($e$), the time of passage through
periastron ($T_0$) and the longitude of periastron ($\omega$).
Three unknowns remain undetermined, the two component masses
($m_1$, $m_2$) and the orbital inclination ($i$). For most spectroscopic
binaries (and most binary pulsars) these quantities are simply not available.

However, there is an equation that links these three unknowns, the mass function:
\begin{equation}
\label{eq:mass_function}
f = \frac{4 \pi^2}{G} \frac{x^3}{P_b^2} = \frac{(m_2 \sin i)^3}{(m_1 + m_2)^2},
\end{equation}
where $G$ is Newton's gravitational constant. This allows an estimate of
one parameter (say, $m_2$) from assumptions for the values of the
other parameters ( $m_1$ and $i$).

Two more equations are needed to solve these 3 unknowns. No other equation
can be obtained from the Keplerian parameters, and generally no more 
equations are readily available. 

\subsection{Binary Pulsars}

The distinguishing feature of binary pulsars is that
we can determine the {\em range} directly
from the measured time of arrival (T.O.A.s) of the radio pulses. Such a
measurement is impossible with spectroscopic binaries.
This happens because the radio pulses (unlike
spectral lines) repeat predictably at a single frequency in the star's own
reference frame (the neutron star's spin frequency).

This is extremely important because it allows for an astounding gain
in precision. As an example, when the orbit of PSR~J2016+1947 was published
\cite{naf03}, the orbit was estimated from variations in the Doppler shift
of the spin frequency of the
pulsar. The projected size of the orbit and the eccentricity determined from
this method were $x = (150.70 \pm 0.07)$ lt-s and $e = 0.00128 \pm 0.00016$.
After the correct rotation count was determined for all observations of
this pulsar, the range of the pulsar relative to the
center of mass of the binary could be measured directly, with a precision of
a few km in each instance. As a result,
we now obtain $x = (150.7730407 \pm 0.0000009)$ lt-s and
$e = 0.001479863 \pm 0.000000016$. This represents a gain in precision of
about $10^5$ and $10^4$ respectively. {\em This is the fundamental reason
why binary pulsars are superior astrophysical tools}.

For most binary pulsars we are still unable to provide any extra
equations to help solve the $m_1$, $m_2$ and $i$ system,
despite the unmatched precision provided by the accurate ranging.

\subsection{Post-Keplerian Effects}

In a few cases, the precision provided by pulsar timing {\em and} the
peculiarities of the system are such that small "post-Newtonian"
deviations from a Keplerian orbit due to the effects of GR become detectable
in the times of arrival of the pulses. These can be parametrized by five quantities,
known as "Post-Keplerian (PK) parameters": the rate of advance of periastron
($\dot{\omega}$), the ``Einstein delay'' $\gamma$ (due to the larger
than average gravitational redshift and special relativistic time dilation near
periastron), the rate of orbital decay due to gravitational radiation
$\dot{P_b}$ and two parameters that characterize the effect of
the gravitational field of the companion on the propagation of the
pulsar's radio signal
(a.k.a. ``Shapiro'' delay): the ``range'' ($r$) and ``shape'' ($s$).
If GR is the correct description of gravity, the PK parameters are given by:
\begin{eqnarray}
\dot{\omega} & = & 3 \left( \frac{P_b}{2 \pi} \right)^{-5/3}
\left(T_{\odot} M \right)^{2/3} (1 - e^2)^{-1} \label{eq:omdot} \\
\gamma & = & e  \left( \frac{P_b}{2 \pi} \right)^{1/3} T_{\odot}^{2/3}m_c (m_p + 2 m_c) \label{eq:gamma}\\
\dot{P_b} & = & - \frac{192 \pi}{5}  \left( \frac{P_b}{2 \pi} \right)^{-5/3}
f(e) T_{\odot}^{5/3} m_p m_c M^{-1/3} \label{eq:pdot} \\
r & = & T_{\odot} m_c \\
s & = & \sin i, \label{eq:s}
\end{eqnarray}
where $m_p$ and $m_c$ are the pulsar and companion masses, $M$ is the total
mass, $T_{\odot} \equiv G M_{\odot}/c^3 = 4.925490947\,\mu$s and
\begin{equation}
f(e) =  \left( 1 + \frac{73}{24}e^2 + \frac{37}{96}e^4  \right)  (1 - e^2)^{-7/2}.
\end{equation}

The main thing to extract from these equations is that they depend on
precisely measurable Keplerian parameters, but also on combinations of $m_p$,
$m_c$ and $i$. Therefore, the measurement of two PK parameters provides,
together with eq. \ref{eq:mass_function}, enough equations to solve for $m_p$,
$m_c$ and $i$.

The measurement of additional PK parameters allows a test of the
self-consistency of general relativity, or at least a verification that
there are no classical contributions due to one of the components having
a finite size. The most famous example of a GR test is that carried out with
the first binary pulsar ever found, PSR~B1913+16.
The system has an eccentric ($e = 0.617$) and compact
($P_b = 7^{\rm h} 45^{\rm m}$) orbit, which means that we can
measure precisely the longitude of periastron ($\omega =
226.57518(4)^\circ$\footnote{The digit in parenthesis represents the
uncertainty in the previous digit.} on 1986 January 14) and how it changes with time
($\dot{\omega} = 4.226607(7)^\circ \rm yr^{-1}$). The eccentricity causes a
non-zero $\gamma$, which becomes measurable because of the fast
precession: $\gamma = 0.004294(1)$s. {\em Assuming} GR, i.e., assuming that
equations \ref{eq:omdot} and \ref{eq:gamma} apply, we obtain for the CSP
$m_p = 1.4408(3) M_{\odot}$ and for the younger unrecycled
(and undetectable) NS $m_c = 1.3873(3) M_{\odot}$ \cite{wt03}.

These masses allow us to predict the other three PK effects for
PSR~B1913+16 using
eqs. \ref{eq:pdot} - \ref{eq:s}, again assuming that GR applies.
The compactness of this system and the long timing baseline allowed the
measurement of one of these PK effects, the orbital decay due to the emission
of gravitational waves:
$\dot{P_b} = -2.4211(14) \times 10^{-12} \rm s s^{-1}$ \cite{wt03}.
This in perfect agreement with the prediction of eq. \ref{eq:pdot}. Apart from
testing the self-consistency of GR, this measurement demonstrated the
existence of gravitational waves in the Universe. For this measurement
the discoverers, Russel Hulse and Joseph Taylor earned the Nobel Prize in
Physics in 1993.

A useful way of visualizing these consistency tests is by drawing a
{\em mass-mass diagram} (Fig. \ref{fig:1903} represents the
constraints derived for PSR~J1903+0327). In such a diagram,
$m_p$ and $m_c$ are the two orthogonal axes. The measurement of each PK
parameter limits the possibilities for $m_p$ and $m_c$ to a small band of
the mass-mass diagram. In the case of DNS systems, where the components
are point masses, all of these bands must meet at a single point, otherwise
GR fails the test. That has not happened yet: Since the discovery of
PSR~B1913+16 similar tests have been made for other systems (for a review
on this vast subject see \cite{sta03}) and GR has passed all of them.
A system discovered more recently, the "double pulsar" (PSR~J0737$-$3039)
\cite{lbk+04}, allows a total of four tests or GR from timing alone \cite{ksm+06}.

\subsection{Neutron Star Masses}

Both PSR~B1913+16 and PSR~J0737$-$3039A are ``centi-second pulsars''. All
precise neutron star mass measurements made to date come from CSPs located
in DNS systems. They range between $1.2489(7) M_{\odot}$
for PSR~J0737$-$3039B, a ``normal'' pulsar that was not recycled at all
\cite{ksm+06} and $1.4408(3) M_{\odot}$ for PSR~B1913+16 \cite{wt03}.
These values are very close to the Chandrasekhar mass ($\sim 1.4 M_{\odot}$),
the upper mass limit for white dwarfs, above which these objects become
gravitationally unstable.

Is such a narrow distribution of masses also observed for MSPs? From our
brief discussion on how they form,
MSPs could in principle be more massive, owing to the much longer accretion
episode that formed them, but until recently no MSPs
had their masses determined to less than 10\% precision and no
precise GR tests have been carried out in MSPs-WD systems.

This state of affairs is somewhat surprising, particularly considering that
the timing precision for MSPs can be one or two orders of magnitude better
than for CSPs. However, as we remarked in the section on the formation of
binary pulsars, systems containing MSPs form with very small orbital
eccentricities ($2 \times 10^{-3}$ to $ 10^{-7}$). As remarked above
for PSR~B1913+16, it is the eccentricity of the
orbit that allows a measurement of $\dot{\omega}$ and $\gamma$, and it is
the combination of these two parameters that generally provides the precise mass
measurements for the NSs in DNS systems. For the circular orbits of
MSP-WD binaries, no such measurements are possible.

So what about the other PK parameters? They don't require an
eccentric orbit to be measured, but $\dot{P_b}$ is greatly amplified
by a) a compact orbit, b) a large orbital eccentricity and
c) a massive companion (eq. \ref{eq:pdot}).
The latter requirements mean that even for MSP-WD systems in compact orbits
(where WD masses are of the order of 0.1-0.2 $M_{\odot}$ and the
eccentricities are always very small) the orbital decay is is much smaller than
in DNS systems. This makes its measurement quite difficult, despite
the much better timing precision of MSPs. Nevertheless, orbital period
decays due to GW emission have been measured and used to estimate MSP
masses (e.g., \cite{nsk08}).

How about the Shapiro delay? This is
a small timing effect (typically, of the order of a few $\mu$s in the T.O.A.s)
which, being proportional to $m_c$, is significantly smaller in MSP-WD systems
than in CSP-NS systems. However, the excellent timing precision of MSPs
more than makes up for the smaller $m_c$. Nevertheless, this effect
seldom provides a precise estimate for the pulsar mass in CSPs or MSPs.

This situation is somewhat frustrating: MSPs could in principle
have higher masses than CSPs and normal pulsars, but despite their excellent
timing precision, no precise mass measurements can be
carried under normal circumstances.  This results in a lack of GR tests made
with MSP-WD systems, which is a great pity: the difference in binding
energies between the components makes MSP-WD systems potentially very
powerful in detecting violations of the strong equivalence
principle (SEP) \cite{sta03}, the fundamental physical basis of GR.

This situation has started to change in recent years, as described below.

\section{Eccentric Binary Millisecond Pulsars in Globular Clusters}
\label{sec:eccentric}

In globular clusters low-mass X-ray binaries are three orders of magnitude
more common than in the Galaxctic disk.
Given the unusual stellar density at the cores of globular clusters,
it is likely the MS stars are captured by old NS lurking near the
centers of these clusters. Evolution of such MS stars leads to accretion
into the NS, i.e., large numbers of X-ray binaries \cite{cla75}. These
eventually become MSPs.

Since 1987, 140 pulsars have been discovered in
globular clusters (GCs)\footnote{For an updated list, see
http://www2.naic.edu/$\sim$pfreire/GCpsr.html.}, 3/4 of the total were
discovered in the last ten years. As predicted, the vast majority of this
pulsar population consists of MSPs, a situation that is very different
than what is observed in the Galaxy.

As in the Galaxy, the X-ray binary phase produces to low-eccentricity
orbits, but in GCs gravitational interactions with passing stars (or even
exchange encounters) can, in a few cases, produce binary MSPs with
highly eccentric
orbits \cite{rh95}. The first of the very eccentric ($e > 0.3$) binary MSPs
to be discovered, PSR~J0514$-$4002A, a 5-ms pulsar located in NGC~1851
has $e = 0.888$ \cite{fgri04}; more so than for any
recycled pulsar in the Galaxy.

Such high eccentricities have already allowed the measurement of
$\dot{\omega}$ for all the eccentric binary MSPs with well-known orbits
(see Table \ref{tab:a}). If the observed $\dot{\omega}$
is entirely due to the effects of GR (if, e.g., there are no classical
contributions due to the finite size of the companion star), then this allows
an estimate of the total mass of the binary $M$ (see eq. \ref{eq:omdot}).
These values are listed in  Table \ref{tab:a}.

As we have seen above, two equations (for the mass function $f$ and
$\dot{\omega}$) are not enough to determine the three unknowns $m_p$, $m_c$
and $i$. In eccentric orbits, we can always measure $\gamma$ and
obtain the third equation needed to solve the system.
However, to be able to do so, we must wait for the orbit to precess:
if there is no precession, the effect on the T.O.A.s
parametrized by $\gamma$ can be
absorbed by a small re-definition of the size of the orbit, $x$. The orbits
of the eccentric MSPs are not as compact as those of the DNS systems for
which we have measurements of $\gamma$; this means that the relativistic
precession is much slower. The end result is that it takes many years
of timing (more than has elapsed since the discovery of the eccentric
binary MSPs in Table \ref{tab:a}) to measure $\gamma$.

Therefore, the $m_p$-$m_c$-$i$ system cannot yet
be solved for any of the eccentric binary MSPs in globular clusters.
Despite this limitation, several important things can be learned from a 
measurement of $\dot{\omega}$ alone, as discussed below.

\subsection{Early Results}

The first measurement of $\dot{\omega}$ for a MSP was made in 2003 for
PSR~J0024$-$7204H, a  3.21-ms pulsar in the globular cluster 47~Tucanae
\cite{mrb+91}. The pulsar is in a mildly eccentric ($e = 0.07$)
binary system with an orbital period of 2.35 days \cite{clf+00}. The
measured periastron advance
($\dot{\omega} = (0.066 \pm 0.001)^\circ \rm yr^{-1}$, \cite{fck+03})
implies, under the assumption of no classical contributions,
$M = (1.61 \pm 0.04) M_{\odot}$.

As mentioned above, this measurement alone is not enough to determine $m_p$,
$m_c$ and $i$. However, we can combine eq. \ref{eq:mass_function} with $M$
(derived from $\dot{\omega}$) to determine a minimum companion mass from
the condition $\sin i \leq 1$:
\begin{equation}
m_c > (f M^2)^{1/3}.
\end{equation}
A maximum pulsar mass can then be derived from $m_p = M - m_c$; the idea
is illustrated graphically in Fig.~\ref{fig:M5B}.
In the case of PSR~J0024$-$7204H, this implies $m_p < 1.52\, M_{\odot}$, i.e.,
the mass of this MSP cannot be much higher than the masses we find in DNS
systems.
The important point about this result (which was not even briefly mentioned
in \cite{fck+03}) is that it shows that at least some NSs can be spun up to
MSP periods with relatively small ($d m < 0.2 M_{\odot}$) amounts of matter.

The discovery of PSR~J1909$-$3744 \cite{jbk+03} in a search of 
intermediate Galactic latitudes provided a similar result.
This 2.9-ms pulsar is interesting because it has the narrowest pulse
profile for any known pulsar ($w_{50} = 43 \mu$s); as a result, it is
one of the most precisely timed MSPs. Furthermore, the orbital
inclination is close to $90^\circ$.
This fortunate combination of characteristics allowed the most precise
measurement of a Shapiro delay to date. The precise
values for $r$ and $s$ (and $f$) provide unambiguous values
for $m_c$, $i$ and $m_p = (1.438 \pm 0.024) M_{\odot}$ \cite{jhb+05}.
Again, this is consistent with the masses measured in DNS systems. Similar
results have been obtained for other MSPs (see Table \ref{tab:a}).

At the start of 2005, it was known that at least some MSPs have masses
similar to the NSs found in DNS systems. Spin-up to MSP periods
can definitely be achieved with small amounts of matter.

\begin{table}
\label{tab:a}
\begin{tabular}{ l c r c c c c c c c c }
\hline
    \tablehead{1}{l}{b}{Name PSR}
  & \tablehead{1}{c}{b}{GC}
  & \tablehead{1}{c}{b}{$P$ (ms)}
  & \tablehead{1}{c}{b}{$P_b$ (days)}
  & \tablehead{1}{c}{b}{$e$}
  & \tablehead{1}{c}{b}{$f/M_{\odot}$}
  & \tablehead{1}{c}{b}{$M/M_{\odot}$ (a)}
  & \tablehead{1}{c}{b}{$M_c/M_{\odot}$}
  & \tablehead{1}{c}{b}{$M_p/M_{\odot}$}
  & \tablehead{1}{c}{b}{Method (b)}
  & \tablehead{1}{c}{b}{Ref. (c)}   \\
\hline
\multicolumn{11}{c}{MSP Mass Measurements} \\

\hline

J0751+1807 & - & 3.47877 & 0.26314 & 0.00000 & 0.0009674 & - & - &
$1.26^{+14}_{-12}$ & $\dot{P_b}$, $s$ & \cite{nsk08} \\

J1911$-$5958A & NGC~6752 & 3.26619 & 0.83711 & $<$0.00001 &
0.002688 & 1.58$^{+0.16}_{-0.10}$ & 0.18(2) & $1.40^{+0.16}_{-0.10}$ & Opt. &
\cite{bkkv06} \\

J1909$-$3744 & - & 2.94711 & 1.53345 & 0.00000 & 0.003122 &
$1.67^{+3}_{-2}$ & 0.2038(22) & $1.438(24)$ & $r,s$ &  \cite{jhb+05} \\

J0437$-$4715 & - & 5.75745 & 5.74105 & 0.00002 & 0.001243 & 2.01(20) &
0.254(14) & 1.76(20) & $r,s$ & \cite{vbs+08} \\

J1903+0327 & - & 2.14991 & 95.1741 & 0.43668 & 0.139607 & 2.88(9) &
1.051(15) & 1.74(4) & $\dot{\omega}$, $s$ &   \cite{crl+08} \\

\hline

\multicolumn{11}{c}{Binary systems with indeterminate orbital inclinations} \\

\hline

J0024$-$7204H & 47~Tucanae & 3.21034 & 2.35770 & 0.07056 & 0.001927 & 1.61(4) &
 $> 0.164$ & $< 1.52$ & $\dot{\omega}$ & \cite{fck+03} \\

J1824$-$2452C & M28 & 4.15828 & 8.07781 & 0.84704 & 0.006553 & 1.616(7) &
 $> 0.260$ & $< 1.367$ & $\dot{\omega}$ &  \cite{brf+08} \\

\hline

J1748$-$2446I & Terzan~5 & 9.57019 & 1.328 & 0.428 & 0.003658 & 2.17(2) &
$> 0.24$ & $< 1.96$  & $\dot{\omega}$ &  \cite{rhs+05} \\

J1748$-$2446J (c) & Terzan~5 & 80.3379 & 1.102 & 0.350 & 0.013066 & 2.20(4) &
$> 0.38$ & $< 1.96$  & $\dot{\omega}$ &  \cite{rhs+05} \\

B1516+02B & M5 & 7.94694 & 6.85845 & 0.13784 & 0.000647 & 2.29(17) &
$> 0.13$ & $< 2.52$  & $\dot{\omega}$ & \cite{fwbh08} \\

J0514$-$4002A (d) & NGC~1851 & 4.99058 & 18.7852 &
0.88798 & 0.145495 & 2.453(14) & 
$> 0.96$ & $< 1.52$ & $\dot{\omega}$ &  \cite{frg07} \\

J1748$-$2021B & NGC~6440 & 16.76013 & 20.5500 & 0.57016 & 0.000227 & 2.91(25) &
$> 0.11$ & $< 3.3$ & $\dot{\omega}$ & \cite{frb+08} \\

\hline

\end{tabular}
\caption{MSP binaries . Notes: a) Binary systems are sorted according to the total
estimated mass $M$. b) Methods are: $\dot{P_b}$ - relativistic orbital
decay, $r,s$ - Shapiro delay ``shape'' and ``range'', ``Opt'' -
optically derived mass ratio, plus mass estimate based on spectrum
of companion, $\dot{\omega}$ - precession of periastron.
c) This pulsar is not technically a MSP, its spin period is longer than
those found in most DNS systems. However, given the similarity of its
orbital parameters to those of Terzan~5~I, we assume that it had a similar
formation history. d) Because of its large companion mass and eccentricity,
this system is thought to have formed in an exchange interaction \cite{fgri04}.}
\end{table}

\subsection{Recent results}

\begin{figure}
  \label{fig:M5B}
  \includegraphics[height=0.35\textheight]{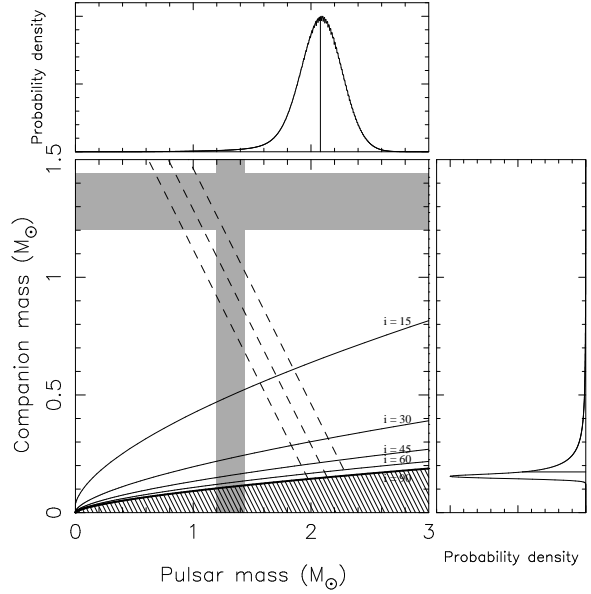}
  \caption{Mass-Mass diagram for PSR~B1516+02B, located in the globular
  cluster M5.  The combination of relativistic $\dot{\omega}$ and the mass
  function provides an unambiguous maximum for $m_p$ and minimum for
  $m_c$. The probability distribution functions on top and left are calculated
  assuming a probability for the orbital inclinations that is constant in $\cos i$
  space. 50\% of the probability lies between the $i = 90^\circ$ and the
  $i = 60^\circ$ lines.}
\end{figure}

In 2005, \cite{rhs+05} discovered 21 new MSPs
in the globular cluster Terzan~5. Nine other MSPs have
been found since, including the fastest-spinning pulsar known
PSR~J1748$-$2426ad, henceforth Ter~5~ad \cite{hrs+06},
making for a total of 33 pulsars in this cluster alone. A total of seventeen
pulsars are members of binary systems; six of these have very eccentric
($e > 0.3$) orbits. This means that Terzan~5 alone
contains half the known population of eccentric binary MSPs.

At the time of publication, only two eccentric binary pulsars
Ter~5~I and J had phase-coherent 
timing solutions (i.e., an unambiguous pulse count for all detected pulses) with
precise measurements of $\dot{\omega}$. Assuming the effect to be relativistic
(a good assumption, because if the companion was extended the orbits of
these two pulsars would circularize in about $10^5$ yr) the
total system masses are $(2.17 \pm 0.02) M_{\odot}$ and
$(2.20 \pm 0.04) M_{\odot}$. The upper mass limits derived for these two
pulsars are very similar, $1.96\,M_{\odot}$.

In principle, this is consistent with the masses being within the range observed
for the NSs in DNS systems. However, it is likely that the pulsar masses are 
significantly higher than $1.44\, M_{\odot}$. The small mass
functions for these systems imply that the companions are likely to
contribute little to the total binary mass.

The argument is probabilistic, and it arises in part
from assuming that there is no
preferred orientation for the binary orbits. For randomly aligned orbits, it is
much more likely that the orbital inclination is close to $90^\circ$
(edge-on orbits) then to $0^\circ$ (face-on orbits). The reason for
this is that for face-on orbits only one possible orientation exists
(orbital plane = plane of the sky), while for edge-on orbits there is
an infinite number of possible orbital planes containing the line of
sight. For a system with no known orbital inclination, the {\em a
  priori} probability of  $i_1 < i < i_2$ is given by $\cos i_1 - \cos i_2$.

Taking this, $\dot{\omega}$ and its uncertainty into account, we can
calculate a probability distribution function (p.d.f.) for the mass of
any MSP and
for the mass of its companion. For Ter~5~I and J, we obtain median values
for the masses above $1.7\,M_{\odot}$ in both cases. Combining the p.d.f.s
for both pulsars, we obtain a 95\% probability that at least one of the NSs has
a mass over $1.68\,M_{\odot}$.

Two other results suggest the possibility of even more massive pulsars.
PSR~B1516+02B is a 7.9-ms binary MSP in a mildly eccentric ($e = 0.14$)
orbit located in the globular cluster M5 \cite{awkp97}. After 19 years
of Arecibo timing the $\dot{\omega}$ for
this system has been finally measured to good precision \cite{fwbh08}.
The total mass of the system is $(2.29 \pm 0.17) M_{\odot}$. This
and the small Keplerian mass function for the companion
imply $m_p = 2.08^{+0.18}_{-0.19}\,M_{\odot}$ (these are
median and 1-$\sigma$ limits). There is a 0.7\% probability that the pulsar
has a mass similar to the NSs found in DNS systems, and there is a 95\%
probability of $m_p > 1.72 M_{\odot}$ (see Fig. \ref{fig:M5B}).

As in the case of Terzan~5~I and J, we have to make the assumption that
the precession of periastron is relativistic. In the case of M5B, this is a
relatively safe assumption: HST archival data shows no companion at the
location of
the pulsar. This implies that it is a small-sized star, either a WD or a
main-sequence star with $m_c < 0.3 M_{\odot}$. In either case, the
contribution of the companion's rotation to the observed $\dot{\omega}$
is likely to be small \cite{fwbh08}.

An even more massive NS might be located in the globular
cluster NGC~6440. PSR~J1748$-$2021B is a 16.7-ms pulsar with an
eccentric ($e = 0.57$) and wide ($P_b = 20.55$ days) orbit with a
low-mass companion.
The relatively good precision of its timing and the high eccentricity
provide a good measurement of $\dot{\omega}$ \cite{frb+08}. Again,
assuming that there are no classical effects on the orbit, the total
mass of the binary is ($2.92 \pm 0.20) M_{\odot}$. Together with the
small mass function measured for this system, this implies an
extraordinary mass for this pulsar,  ($2.74 \pm 0.20) M_{\odot}$,
with a mere 1\% probability of $m_p < 2.0 M_{\odot}$. Until the
appropriate optical studies are carried out, this result is not as secure
as in the case of M5B.

\subsection{MSP Mass Distribution}

\begin{figure}
  \label{fig:psr_masses}
  \includegraphics[height=0.30\textheight]{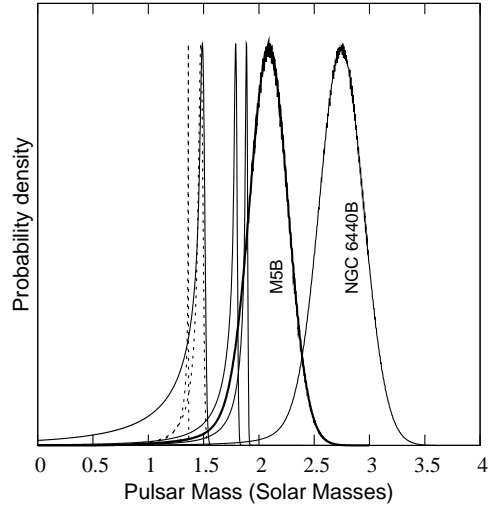}
  \caption{Probability distribution functions
(pdfs) for the eccentric MSPs binaries in GCs.
The mass pdfs of the MSPs in the less
massive binaries (those with $M < 2 M_{\odot}$) are represented by the
dashed curves. The distribution of masses is much broader than is
found for the components of DNS systems. Out of a total
of seven systems four seem to be significantly more massive than the most
massive NS in DNS systems, PSR~B1913+16.}
\end{figure}

The timing results for these MSPs do not prove that there are neutron stars
with masses significantly in excess of $1.44 M_{\odot}$. It is still possible
(but highly unlikely) that there are classical contributions to the observed
$\dot{\omega}$ for most of them. They are, however, highly suggestive:
plotting the p.d.f.s. and taking them at face value, we see that the mass
distribution for MSPs is much broader that that observed for CSPs and normal
NSs in DNS systems (see Fig. \ref{fig:psr_masses}).
About half of the eccentric MSPs in globular clusters seem to have large
($> 1.6\, M_{\odot}$) masses.

It appears therefore that, although some MSPs were spun up with
relatively modest amounts of matter, that was not the case for all
MSPs. The reasons for this somewhat surprising result are not known at
present.

In the Galaxy, the situation appeared to be different until
2008, with PSR~J0751+1807 and PSR~J1909$-$3744 indicating relatively
low masses and other MSPs yielding inconclusive results.
The situation has now changed for Galactic MSPs as well with the
mass measurement of PSR~J0437$-$4715 \cite{vbs+08} and in particular
the discovery of PSR~J1903+0327 \cite{crl+08}.

\section{PSR~J1903+0327}

\begin{figure}
  \label{fig:1903}
  \includegraphics[height=0.43\textheight]{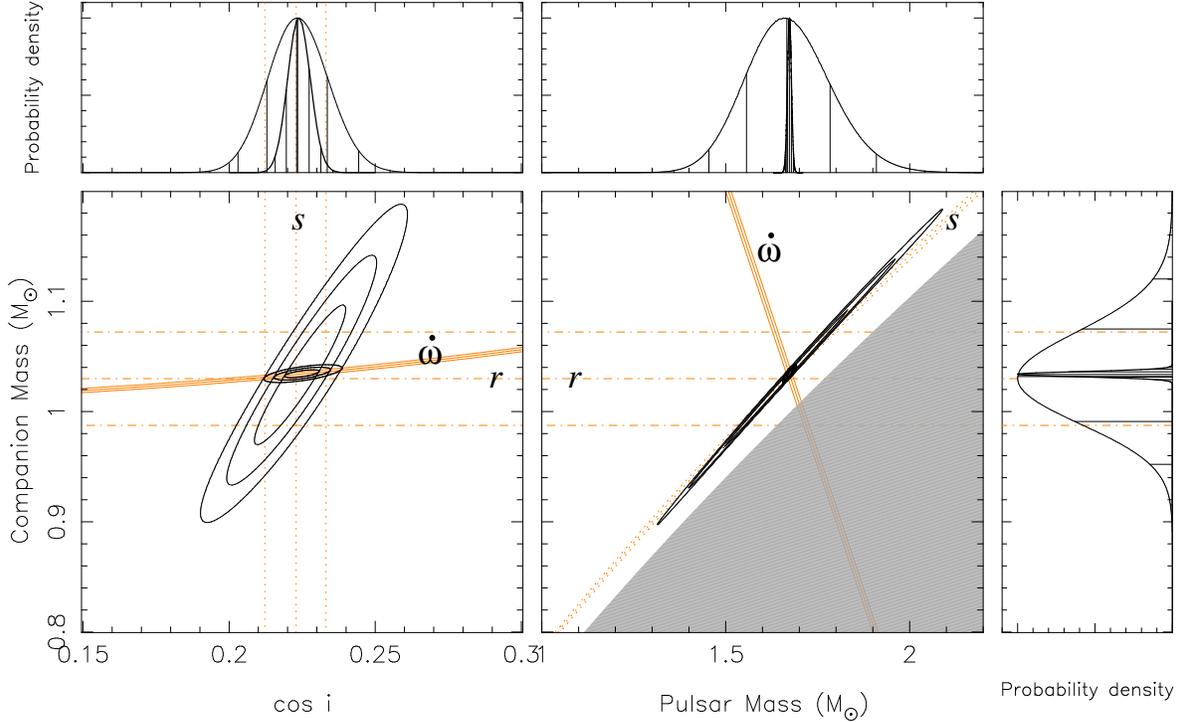}
  \caption{$\cos i$-$m_c$ and $m_p$-$m_c$ diagrams for PSR~J1903+0327.
The contour levels enclose 99.7\%, 95.44\% and 68.3\% of all
probability. The light contour levels are derived from a $\chi^2$ map
calculated from $r$ and $s$ only; the heavy contour levels are
calculated assuming that the observed $\dot{\omega}$ is relativistic.
{\em Top} and {\em Right}: projected 1-D p.d.f.s for $\cos i$, $m_p$
and $m_c$. The pulsar and companion p.d.f.s are much narrower when we
take the $\dot{\omega}$ into account, but entirely within the
regions predicted using $r$ and $s$ alone. Orange lines: regions of the
diagrams consistent with the PK parameters and their 1-$\sigma$
uncertainties estimated by {\sc tempo2}.}
\end{figure}

PSR~J1903+0327 was a the first MSP found in the ongoing ALFA
pulsar survey \cite{cfl+06}. It is a 2.15-ms pulsar in a 95-day orbit
with a $\sim 1
M_{\odot}$ main-sequence star companion. That is very unusual, but
even more unusual is the orbital eccentricity of the system: $e =
0.44$. The standard evolutionary scenarios briefly described at the
start of this review cannot explain the formation of such a system;
this is one of the reasons why this MSP is so interesting. As in the
case of the isolated MSPs in the Galactic disk
the formation of this system is still not well understood \cite{crl+08}.
As an example, it is possible that PSR~J1903+0327 formed in the
same way as the isolated MSPs, i.e., by somehow eliminating its mass
donor, which was much closer to the
pulsar than its present MS star companion. The latter object has been
there from the start, formerly as the outer element of a hierarchical triple;
it has never interacted significantly with the MSP.

The other reason why this binary MSP is so interesting is that its
unusual characteristics mean that 3 PK parameters can be measured precisely.
The eccentricity allows
a precise measurement of $\dot{\omega}$ and the large companion mass
($m_c \sim 1 M_{\odot}$) allows a measurement of $r$ and $s$.
Combining $\dot{\omega}$ and $s$ as measured at the end of
2007 \cite{crl+08} obtained $m_p = (1.74 \pm 0.04) M_{\odot}$; with the
ever-present qualifier that we are assuming $\dot{\omega}$ to be relativistic.
At this time only a small fraction of the orbit had been measured at high timing
precision (2.2 GHz) with Arecibo, and $r$ could not be measured precisely.

To measure $\dot{\omega}$ and $s$ more precisely and verify whether
$\dot{\omega}$ is relativistic, we have started a dense timing campaign
with Arecibo. The idea is to measure $r$ precisely and see if it is
consistent with the companion mass we derive from $\dot{\omega}$ and $s$.

When the first 2.2-GHz orbit was completed with Arecibo, the pulsar
mass estimate decreased to $(1.67 \pm 0.01 ){\rm M}_{\odot}$, but it
has been stable at that level the last 18 months.
The companion mass derived from the
latest values of $\dot{\omega}$ and $s$ is $(1.028 \pm 0.004){\rm M}_{\odot}$.
The latest value measured for $r$ is $(1.03 \pm 0.04) M_{\odot}$. This
agreement apparently
confirms the assumption that the $\dot{\omega}$ is relativistic. However,
because of the lower precision in the measurement of $r$, it is impossible
to exclude small contributions to $\dot{\omega}$. At the moment, if we
calculate the p.d.f. for the mass of PSR~J1903+0327 based on $r$ and $s$
alone (i.e., assuming nothing about $\dot{\omega}$), we obtain
$m_p = (1.67 \pm 0.11) M_{\odot}$ and a 98.4\%
probability that the mass is above $1.44 M_{\odot}$ (see Fig. \ref{fig:1903}).
The precision of $r$ is still improving significantly with continued timing.

The mass value derived for PSR~J1903+0327 from $\dot{\omega}$ and
$s$ is the most precise MSP mass ever measured. The likely contribution
from other effects to $\dot{\omega}$ is now being evaluated, but it is clear
by now that is is likely to be small. After more than 30 years of searches,
this is the first precisely measured NS mass larger than that
of PSR~B1913+16 and the first that is incompatible (and significantly above)
the Chandrasekhar mass. This pulsar proves that accretion can significantly
increase the mass of MSPs compared to the NSs in DNS systems, i.e., the
mass distribution for MSPs is definitely wider than the NS mass distribution
in DNSs.

\section{Studies of Super-dense Matter}

\begin{figure}
  \label{fig:EOS}
  \includegraphics[height=0.55\textheight]{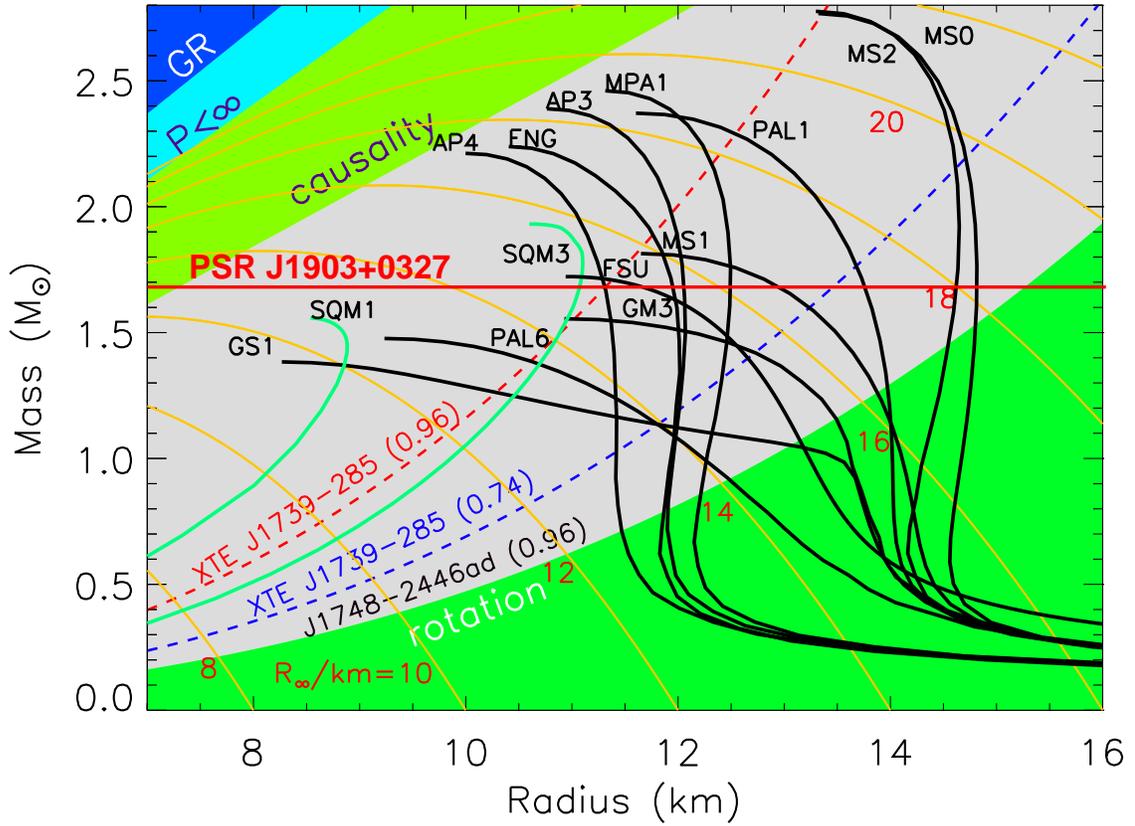}
  \caption{Mass-Radius relation for neutron stars. Each
black curve represents a family of neutron stars masses and
radii according to a given
equation of state. The region bounded by the Schwarzschild condition
$R < 2GM/c^2$ is excluded by general relativity, and that bounded by
$R < 3GM/c^2$ (labeled ``causality'') is excluded by requiring the
speed of sound inside the
star to be smaller than the speed of light. The mass-shedding limit
for the fastest spinning radio pulsar (PSR J1748$-$2446ad, with spin
frequency 716 Hz) is labeled ``rotation''; points in the green region
below this line are not allowed for that particular pulsar.  Stricter
constraints may arise from X-ray sources like
XTE~J1739$-$285 (dashed curves calculated under different neutron
star models) if their spin frequencies are confirmed 
to be higher than that of PSR J748$-$2446ad, potentially excluding some
equations of state (such as GM3) which lie almost entirely below the
rotation curve. A recent, precise millisecond pulsar (MSP) mass
measurement (for PSR~J1903+0327) excludes the ``softest'' EOSs (red
horizontal line). Adapted from \cite{lp07}.}
\end{figure}

Why is a precise measurement of MSP masses so important?
First, because of accretion, they can be more massive than NSs in
DNS systems. If they are, then they test our models of how matter
behaves at the center of NSs.

Given their small size ($R\,\sim\,10\,$km) and large mass ($M \,
\sim \, 1.4 M_{\odot}$), NSs contain some of the densest
matter in the Universe: the core can have densities of
several times
10$^{14} g cm^{-3}$. They are therefore unique astrophysical laboratories 
for testing theories of nuclear matter under high
pressures and densities well in excess of that of the atomic nucleus. 

Because matter at such densities are not readily available on Earth, 
its microscopic composition and the relation between
macroscopic quantities like pressure and density (the equation of state,
or EOS) are essentially unknown.
As Fig.~\ref{fig:EOS} shows, there are large variations
in predicted radii and maximum masses for different candidate
EOSs. These reflect basic uncertainties about
the microscopic behavior and composition of matter at and above nuclear
density. Some EOSs,
such as GS1, assume that large percentages of matter are in exotic states
(hyperons, deconfined quark matter), any of which produces a decrease
in the nucleonic (protonic and neutronic) degeneracy pressure at any
given density, since a larger variety of particles are present.
This means that matter is relatively more compressible; this
results in smaller maximum stellar masses.
Other EOSs, such as MSO, assume a larger fraction of nucleons and
have higher pressures for any given density.  This means that they
predict matter to be relatively incompressible and result in larger
neutron star mass limits.

As we can see in Fig. \ref{fig:EOS}, the mass measured for
PSR~J1903+0327 is higher than the maximum possible mass
predicted by some EOS models still being considered
in the literature. Such models are excluded by the
PSR~J1903+0327 mass measurement.

The mass measurement for PSR~J1903+0327, being the largest known
with certainty, also sets a quantitative 
upper  limit to its central density, which then cannot be exceeded in 
any neutron star of lower mass.  As successively more massive neutron stars are 
observed, the limiting central density gets smaller.  Approximately,
the limit is $\rho_{max}\le36\rho_s({\rm M}_\odot/M_{max})^2$ 
where $\rho_s=2/7\times10^{14}$ g cm$^{-3}$ is the nuclear 
saturation density.  A large enough observed mass ($2 M_{\odot}$) could rule
out the  appearance of exotic phases in neutron stars, such as deconfined 
quarks \citep{lp05}; these are indicated by the SQM lines in Fig. \ref{fig:EOS}.


\section{Prospects}

In Terzan 5, we now have measurements of $\dot{\omega}$ for a total of seven
binary systems. Two more such measurements have been obtained
for two new eccentric binary pulsars in M28, PSR~J1824$-$2452C and D
\cite{brf+08}. The increased number of mass p.d.f.s will assist the statistical
studies of the MSP mass distribution.

For the eccentric binary MSP with the shortest orbital period,
Ter~5~I, $\gamma$ is now becoming detectable. This will yield
a precise MSP mass measurement for a second system (after PSR~J1903+0327)
where we expect the pulsar to be massive. Whether such a high mass will be
confirmed or not remains to be seen.

Meanwhile, HST time has already
been allocated for the study of the environs of these candidate super-massive
pulsars. The detection (or not) of their companions will be important to
address the issue of whether the observed $\dot{\omega}$ is purely
relativistic or not. In the case of M5B, the non-detection of the companion
in archival images seems to indicate that this is indeed the case, but
that is not as clear in the case of NGC~6440B.

For the latter binary,
our simulations indicate that it will take about 15 years to
determine $\gamma$ with any useful precision and determine the mass of the
pulsar unambiguously, but the potential rewards are immense, particularly for
the study of the EOS: a mass well in excess of $2 M_{\odot}$ would
exclude most of the EOSs now being proposed.
We expect that advances in instrumentation, with
resultant improvements in timing precision, will lead to significant results
well before that.

The prospects for the study of PSR~J1903+0327 are very bright, particularly
at optical wavelenghts (no pun intended here!). Measurements of the
spectral line widths of the
companion might address once and for all the issue of whether the companion
is rotating fast or not and whether it can contribute to the observed
advance of periastron. If not, then we can consider that the combination
of 3 PK parameters ($\omega, r, s$) provides one test of GR.
Measurements of the spectral line shifts will also determine the mass ratio
precisely, providing one extra constraint in the system, i.e., one extra
test of GR.

Finally, new searches are likely to find several more eccentric binary
MSPs, particularly in GCs. 



\begin{theacknowledgments}
PCCF acknowledges support from a WVEPSCoR research challenge
grant held by the WVU Center for Astrophysics.
\end{theacknowledgments}



\end{document}